\documentclass[11pt,a4paper]{article}

\usepackage{jheppub}

\usepackage{amssymb,amsmath}
\usepackage{multicol,bbm}
\usepackage{bm}
\usepackage{multirow}
\usepackage{comment}
\usepackage{mathrsfs}
\usepackage{graphicx}
\usepackage{slashed}
\usepackage{epstopdf}
\usepackage[utf8]{inputenc}
\usepackage[dvipsnames]{xcolor}

\newcommand{\TT}[1]{{\bf\color{magenta} Tim: #1}}

\def\beq{\begin{equation}}
\def\eeq#1{\label{#1}\end{equation}}
\def\eeqn{\end{equation}}
\def\beqa{\begin{eqnarray}}
\def\eeqa#1{\label{#1}\end{eqnarray}}
\def\eeqan{\end{eqnarray}}

\def\stacksymbols #1#2#3#4{\def\theguybelow{#2}
	\def\vp{\lower#3pt}
	\def\sp{\baselineskip0pt\lineskip#4pt}
	\mathrel{\mathpalette\intermediary#1}}

\def\intermediary#1#2{\vp\vbox{\sp
		\everycr={}\tabskip0pt
		\halign{$\mathsurround0pt#1\hfil##\hfil$\crcr#2\crcr
			\theguybelow\crcr}}}

\def\gsim{\stacksymbols{>}{\sim}{2.5}{.2}}

\newcommand{\TeV}{~\textrm{TeV}}
\newcommand{\GeV}{~\textrm{GeV}}

\newcommand{\tab}[1]{Table~\ref{tab:#1}}

\newcommand{\bea}{\begin{eqnarray}}
\newcommand{\eea}{\end{eqnarray}}

\preprint{UCI-HEP-TR-2018-23}

\title{Six Top Messages of New Physics at the LHC}

\author[a]{Huayong Han,}
\author[b,c,d]{Li Huang,}
\author[b,d]{Teng Ma,}
\author[b,c,e,f]{Jing Shu,}
\author[g]{Tim M.P. Tait,}
\author[h]{and Yongcheng Wu}

\emailAdd{han@itp.ac.cn, huangli@itp.ac.cn, mat@itp.ac.cn, jshu@itp.ac.cn, ttait@uci.edu, ycwu@physics.carleton.ca}

\affiliation[a]{Guizhou Key Laboratory in Physics and Related Areas, Guizhou University of Finance and Economics, Guiyang 550025, China}
\affiliation[b]{CAS Key Laboratory of Theoretical Physics, Institute of Theoretical Physics,
Chinese Academy of Sciences, Beijing 100190, China}
\affiliation[c]{School of Physical Sciences, University of Chinese Academy of Sciences, Beijing 100049, P. R. China}
\affiliation[d]{Laboratory for Elementary Particle Physics, Cornell University, Ithaca, NY 14853, USA}
\affiliation[e]{CAS Center for Excellence in Particle Physics, Beijing 100049, China}
\affiliation[f]{Center for High Energy Physics, Peking University, Beijing 100871, China}
\affiliation[g]{Department of Physics and Astronomy, University of California, Irvine, CA 92697 USA}
\affiliation[h]{Ottawa-Carleton Institute for Physics, Carleton University, 1125 Colonel By Drive, Ottawa, Ontario K1S 5B6, Canada}

\abstract{
Six top signatures provide a novel probe of new physics.  We discuss production of six top quarks as the decay products
of a pair of top partners in the setting of a composite Higgs model, and argue that the six top signal may generically provide
one of the first final states to show a discrepancy.  We construct an analysis based on quantities such as
$H_T$ and the numbers of jets which are tagged as boosted tops, $W$s, or containing $b$-tags, and show that
the LHC with 3~ab$^{-1}$ can discover top partners with masses up to around 2.5 TeV in the six top signature.
}

\begin{document}

\maketitle	

\section{Introduction}
\label{sec:introduction}

The Large Hadron Collider (LHC), with its unparalleled energy and high luminosity, 
will definitively explore the physics at the TeV scale. 
The discovery of Higgs boson at the LHC is a triumph of the Standard Model (SM), however, the Naturalness problem associated with the self-energy of the Higgs particle argues that it is likely that there is new physics around the TeV 
scale~\cite{Feng:2013pwa,Giudice:2013nak,Altarelli:2013lla,Farina:2013mla,deGouvea:2014xba,Csaki:2018hyw,Chen:2017dwb}. 
Various new physics models addressing this problem have been proposed, such as Supersymmetry (SUSY), little Higgs, Composite Higgs etc. 
Deep investigation of the naturalness problem may reveal new details underlying the physics of the 
electroweak symmetry breaking (EWSB) and could also provide the evidence of new physics.

Beside the Higgs, the top quark is central to arguments concerning
naturalness, since it has the largest mass of the SM fermions, and hence the largest coupling to the Higgs. 
For this reason,
partners of the top quark are ubiquitous in models of new physics at the weak scale, and their production
often results in multi-top signatures at the LHC, leading to many interesting phenomena.
The four top final state has been previously 
investigated~\cite{Lillie:2007hd,Pomarol:2008bh,Chen:2008hh,Kumar:2009vs,Gregoire:2011ka} 
and is starting to be visible in experimental analysis \cite{Sirunyan:2017roi,Aaboud:2018jsj}. 
However, even more tops in the final state naturally occur under simple assumptions and provides a spectacular collider
signature and a complementary method to search for new physics. 

In this paper, we systematically investigate the phenomenology of six-top final states in a simplified model
inspired by a composite Higgs scenario.
We estimate the sensitivity of the LHC to six-top final states for channels with 
different number of charged leptons, and the upper limit on the top partner branch ratio into $t \bar{t} t$ are
obtained in the case that no signal is observed with 3~ab$^{-1}$ of integrated luminosity.
We also discuss the extraction of the top partner mass.
It should be stressed that six-top final states occur in many other  models of new physics, 
and our general analysis framework can be applied to those cases with simple adjustments.

The paper is organized as follows. In Section~\ref{sec:model}, we introduce a simplified composite Higgs model 
which inspires our analysis and in Section~\ref{sec:signature} discuss general features of the six top
signature and current LHC constraints. 
The analysis strategy of LHC data are described in Section~\ref{sec:LHCPheno}. 
We reserve Section~\ref{sec:conclusion} for our conclusions.

\section{Six Tops from a General Composite Higgs Model}
\label{sec:model}

Generally, composite Higgs models with a simple UV completion 
(such as $SU(4)/Sp(4)$~\cite{Csaki:2017jby,Ryttov:2008xe,Galloway:2010bp} or the isomorphic coset space $SO(6)/SO(5)$~\cite{Gripaios:2009pe,Frigerio:2012uc,Serra:2017poj}  and $SU(4)\times SU(4)/SU(4)$~\cite{Ma:2015gra,Ma:2017vzm,Cacciapaglia:2018avr}), 
contain a singlet scalar pseudo-Nambu-Goldstone boson (pNGB) field $s$ 
corresponding to a broken $U(1)_s$ global symmetry. 
This pNGB can decay into di-bosons through Wess-Zumino-Witten (WZW) terms via fermion loops.
In theories with partial compositeness, $s$ can also decay into fermion pairs through the 
elementary-composite mixing terms between the SM fermions and the composite top partners $t'$. 
Since the decay into dibosons are effectively at loop level, and the large top mass implies in such theories that the top partners
predominantly mix with the SM top, $s$ generically
decays into a top pair with very close to $100\%$ branch ratio (BR). 
The same large mixing generically implies that, provided the mass of the $s$ is not too large, the top partners
themselves decay into $s$ and top with a significant BR.  As a result, a single top partner
typically undergoes the decay chain,
\begin{eqnarray}
t' \rightarrow t s \rightarrow t \bar{t} t,
\end{eqnarray}
and an event originating from pair production of the top partners results in a six top final state
(see \autoref{fig:pairCS} left panel):
\begin{eqnarray}
p\ p \to \bar{t'}\ t' \to \bar{t}\ s\ t\ s \to \bar{t}\ t\ \bar{t}\ t\ \bar{t}\ t.
\end{eqnarray}


We work with an effective Lagrangian capturing the essential features of the interactions between top partners and $s$. 
Requiring that the singlet $s$ renormalizably couples to the top and its partner, the vector-like top partners must 
either be electroweak singlets ($T=t'$) or doublets ($\psi=(t',b')$) with hypercharge $Y=1/6$. 
In the first (singlet) case, the effective Lagrangian reads
\begin{align}
\mathscr{L} &= \bar{T} (i \slashed{D} - m_{t'})T  +\frac{1}{2}\partial_\mu s \partial^\mu s - \frac{1}{2} m_s^2 s^2 
- \lambda ~s \bar{T}_L t_R-\lambda_1 \bar{q}_L H T_R-\lambda_2 \bar{T}_L t_R  + h.c.
\end{align}
And the doublet case is described by
\begin{align}
\mathscr{L} &= \bar{\psi} (i \slashed{D} - m_{t'})\psi +\frac{1}{2}\partial_\mu s \partial^\mu s - \frac{1}{2}m_s^2 s^2 
- \lambda ~s \bar{\psi}_R q_L-\lambda_1 \bar{\psi}_L H t_R-\lambda_2 \bar{\psi}_R q_L  + h.c.
\end{align}
Here $H$ is the SM Higgs doublet field, $D_\mu$ is the appropriate covariant derivative,
$m_{t'}$ and $m_{s}$ are the masses for top-partner and $s$ respectively, 
and $\lambda_i$ are coupling constants.  
We work in the limit where the coupling $\lambda$ is much larger than $\lambda_{1,2}$ or the electroweak coupling, 
such that the top-partner decays are predominantly into top and $s$ with almost 100\% BR, but is small enough that the
width of the top partner remains relatively narrow.
In this limit, the relevant parameters are the top partner and scalar masses, with mild dependence on the 
strength of the interactions.
In the more general case where the top partners have appreciable decays into other channels, 
our results can be rescaled with the corresponding BR and continue to apply.

\section{Top Partner Pair Production and Signatures}
\label{sec:signature}

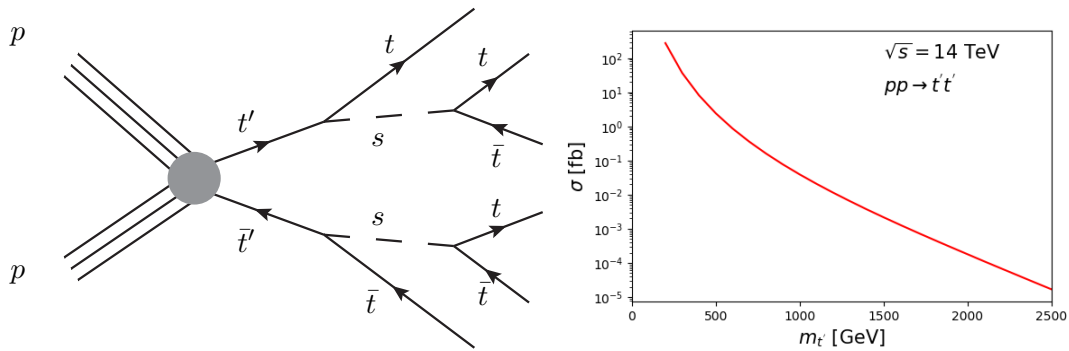
\begin{figure}[!t]
\centering
\includegraphics[width=0.45\textwidth]{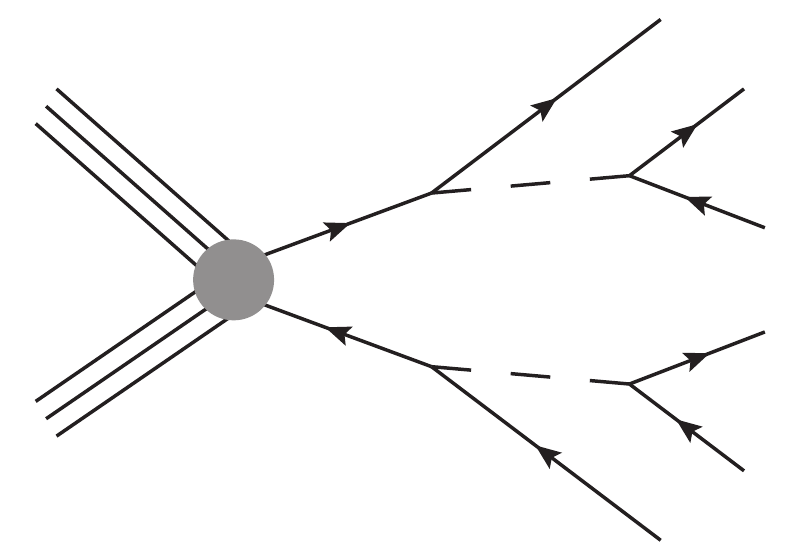}
\put(-205, 30){$p$}
\put(-205, 120){$p$}
\put(-120, 85){$t'$}
\put(-120, 40){$\bar{t}'$}
\put(-65, 115){$t$}
\put(-72, 17){$\bar{t}$}
\put(-70, 80){$s$}
\put(-70, 50){$s$}
\put(-30, 110){$t$}
\put(-30, 20){$\bar{t}$}
\put(-25, 72){$\bar{t}$}
\put(-25, 53){$t$}
\includegraphics[width=0.45\textwidth]{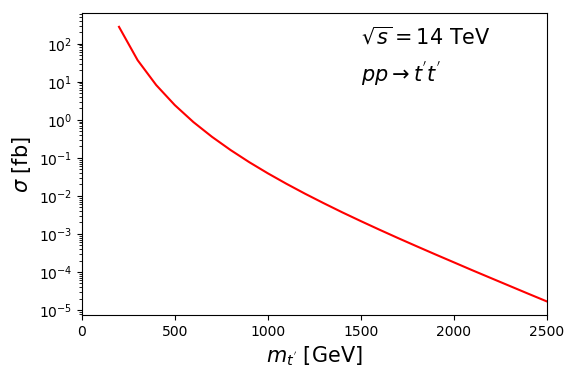}
\caption{Left: Representative Feynman diagram for a 6 top final state through top-partner pair production. 
Right: The cross section for top-partner pair production at the LHC with $\sqrt{s}=14$ TeV as function of top-partner mass.}
\label{fig:pairCS}
\end{figure}

For modest mixing, the dominant top partner production mechanism at the LHC is production of
a $t' \bar{t}'$ pair through the strong force of which the rate only depends on the partner mass and the strong coupling.  The rate at the LHC operating at $\sqrt{s} = 14$~TeV as a function of the top
partner mass is shown in the right panel of \autoref{fig:pairCS}.


As with other multi-top final states, it is convenient to classify six top final states based on the decay modes of the $W$
bosons.  Leptonic decay modes allow for up to six very energetic charged leptons
($\ell = e, \mu$) in the final state.  
In Table.~\ref{tab:signal_channel}, we list the channels containing up to three isolated
charged leptons along with their corresponding branching
ratios and the primary SM backgrounds leading to topologies similar to a six top final state. 
Final states with four or more charged leptons are not considered, as the BR for these channels is highly suppressed.
While several of these channels have previously been analyzed at the LHC~\cite{Aad:2016tuk,ATLAS-CONF-2016-013,Aaboud:2017dmy,Aaboud:2018zeb,Sirunyan:2017lae,Khachatryan:2017qgo}, the focus was on a different production mechanism, and thus not optimized to extract a six top final state.
A six top final state also allows for the new, not previously analyzed, signatures such as three same-sign charged leptons.

In addition to channels with various numbers of leptons, there are several other 
generic features which commonly appear in the six top signature, including:
\begin{itemize}
\item Large $H_T \equiv \sum_i|p_T^i|$ (where the index runs over all visible final state particles),
typically $\gsim 2000$ GeV for the range of $m_{t^{'}}$ under consideration;
\item Boosted top jets which may appear as fat jets in the detector;
\item High multiplicity of bottom-flavored and/or light jets.
\end{itemize}

\begin{table}[!t]
	\centering
	\resizebox*{\textwidth}{!}{
	\begin{tabular}{|c|c|c|c|}
		\hline
		\hline
		Channel &  ~~Branching Fraction~~ & Event Fraction & ~~SM Backgrounds~~\\
		 &  (Truth level) & Reconstructed (1.5 TeV) & \\
		\hline\hline
		1 Lepton & 17.82\% &38.65\% & $t\bar{t}+n$-jets  \\
		\cline{1-3}
		2 Opposite-Sign Leptons &  8.46\% &9.50\% & $t\bar{t}t\bar{t}$ \\
		\cline{1-3}
		2 Same-Sign Leptons & 5.36\% &6.51\% & $t\bar{t}Z+n$-jets\\
		\cline{1-3}
		3 Mixed-Sign Leptons &  5.64\% &3.67\% & $t\bar{t}W+n$-jets \\
		\cline{1-3}
		3 Same-Sign Leptons&  0.60\% &0.71\% & \\
		\hline
		\hline
	\end{tabular}}
	\caption{Analysis channels arising from six-top final states with corresponding branch fraction, organized according to the number of leptons in the final state. The events fraction including possible mixing between different channels when considering mis-identification and detector effects for $m_{t'}=1500$ GeV are listed in the third column. Note that the around 1\% lepton fake rate from the jets which results in more leptons due to the large multiplicity of the jets in each events. 
	Dominant SM backgrounds are also listed in the last column. 
    }
	\label{tab:signal_channel}
\end{table}

\subsection{Current Constraints}

\begin{figure}[!t]
\centering
\includegraphics[width=0.45\textwidth]{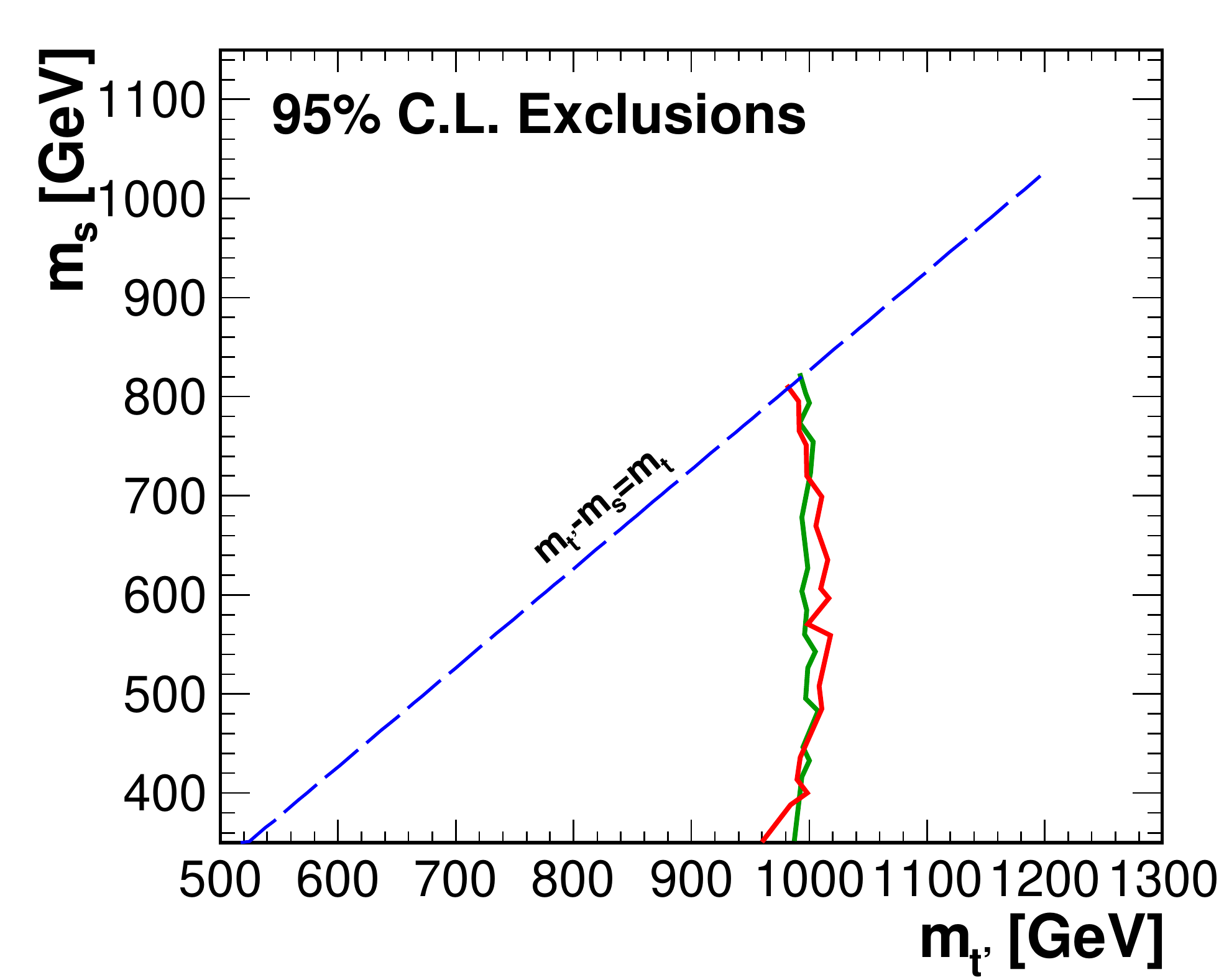}
\includegraphics[width=0.45\textwidth]{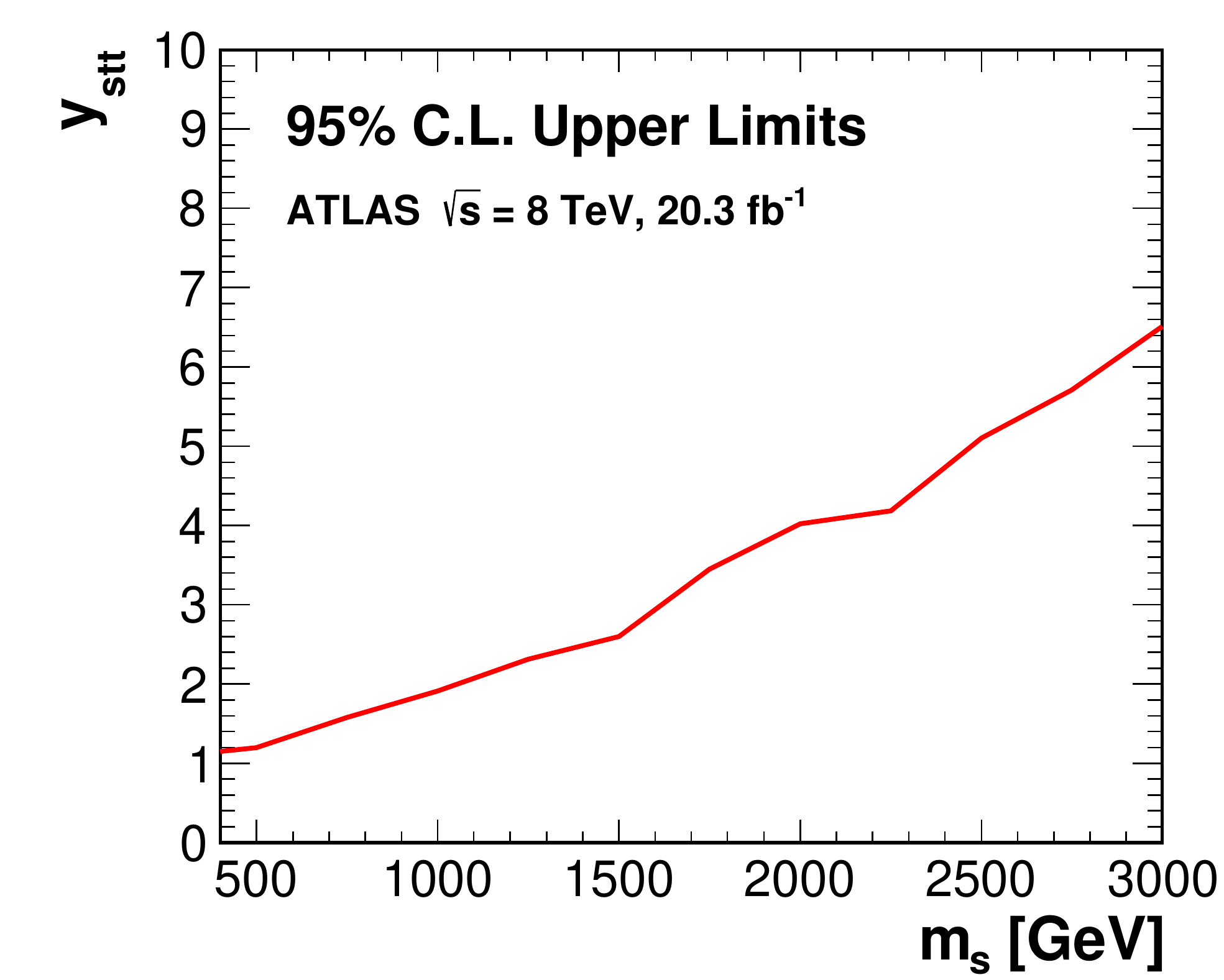}
\caption{Left: Current constraints on the $m_s$-$m_{t'}$ plane from the LHC as implemented in CheckMATE. 
The red line shows the boundary of the excluded region based on the analysis of Ref~\cite{Aad:2016tuk}, 
and the green line corresponds to Ref~\cite{ATLAS-CONF-2016-013}. Right: Upper limit on the $s$-$t$-$\bar{t}$ coupling strength as a function of $m_s$ from LHC searches for top pair production through scalar resonance~\cite{Aad:2015fna}.
}
\label{fig:currentconstraints}
\end{figure}

Most searches for top partners at the LHC have considered missing transverse momentum signatures  (based on SUSY
searches \cite{Aaboud:2017nfd,Aaboud:2017ayj,Aaboud:2017aeu}) which occur in theories in which the top partner is
connected to a dark matter candidate.  These searches exclude scalar top partners with masses up to $900-1000$~GeV,
depending on the mass of the dark matter candidate.
We evaluate the constraints from visible signatures using {\tt CheckMATE}~\cite{Drees:2013wra}, the results are shown in the $m_{t'}$-$m_{s}$ plane in the left panel of \autoref{fig:currentconstraints}. The most stringent constraints are coming from multi-lepton (red line)~\cite{Aad:2016tuk} 
and multi top quarks searches (green line)~\cite{ATLAS-CONF-2016-013}.  
These constraints exclude
cross section $\sigma(pp\to t'\bar{t'}) < 28.63$ fb 
at the $95\%$ C.L. for $\sqrt{s} = 13$ TeV, corresponding to top partner masses up to nearly 1 TeV.

There is also the possibility to directly produce the $s$ from gluon fusion, which results in a $t \bar{t}$ final state
whose invariant mass is resonantly enhanced at $m_s$.
In the right panel of~\autoref{fig:currentconstraints}, we show the observational upper limit 
derived from 8 TeV LHC search for resonant top pair production~\cite{Aad:2015fna} on the
$s$-$t$-$\bar{t}$ coupling strength as a function of the $s$ mass. Note that, here we only present the constraints from 8 TeV analysis. New 13 TeV searches~\cite{Aaboud:2018mjh} will definitely improve the sensitivity. However, the detailed reanalysis of the 13 TeV result in our scheme is beyond our scope, we leave this for future works.

\section{Identifying Six Top Events at the LHC}
\label{sec:LHCPheno}

We divide our analysis into channels with 1, 2 or 3 isolated leptons (1, 2, 3-$\ell$) in the final state. 
The 2- and 3-lepton channels are further divided according to the charges of the isolated leptons. Hence in total, we have five different channels: 1-lepton, 2 opposite sign leptons (2-os$\ell$), 2 same sign leptons (2-ss$\ell$), 3 mixed sign leptons (3-ms$\ell$) and 3 same sign leptons (3-ss$\ell$). These channels are by definition orthogonal to each other, such that 
a direct combination is straightforward.

\subsection{Simulation and Event Reconstruction}

We simulate signal and background events for the LHC running at $\sqrt{s} = 14$~TeV.  Events are generated at the parton
level via the {\tt MadGraph5} package~\cite{Alwall:2014hca}, using {\tt CTEQ6L} 
parton distribution functions (PDFs)~\cite{Nadolsky:2008zw}. 
Resonances are decayed
either via {\tt MadSpin}~\cite{Artoisenet:2012st} for top quarks and $W$ bosons, or {\tt PYTHIA8}~\cite{Sjostrand:2014zea}
for the top partners.  Parton level events are then passed to {\tt PYTHIA8}
for initial state radiation, showering and hadronization. 
The detector reconstruction is simulated by {\tt Delphes}~\cite{deFavereau:2013fsa} using the default CMS configuration with modified lepton isolation and b-tagging efficiency (described below).
Selection cuts are imposed through the {\tt ROOT} framework via the PyROOT interface,
with {\tt FastJet}~\cite{Cacciari:2011ma}
providing further jet reconstruction and clustering analysis.

The signal process is generated as  $pp \rightarrow t' \bar{t}'$ for the set of top partner masses
$m_{t'} =$ 1.0, 1.3, 1.5, 1.8, 2.0 and 2.5$\TeV$.  
As mentioned above, {\tt PYTHIA8} decays the top partners into top quarks
via $t' \rightarrow t s \rightarrow t \bar{t} t$, with an assumed $100\%$ branching ratio.  This process loses information
regarding spin correlations, and thus we do not explore related observables in this analysis.
For each choice of $m_{t'}$, we fix the singlet mass to be $m_s = m_{t^{'}}-500\GeV$.  While 
this choice is not general, our analysis does
not rely on any selection related to this choice, and so we expect the derived efficiencies to be roughly independent of $m_s$.
However, the kinematic endpoints $m_{t^{'}}\approx m_s + m_t$ or $m_s \approx 2m_t$ produce unusually soft top quarks,
which could impact the distribution of events containing top quarks or $W$ bosons reconstructing as fat jets.  We minimize
the impact by restricting ourselves to softer requirements on the corresponding variables, but it would be worthwhile to explore
this region of parameter space in more detail.

The background processes are generated as:
\begin{itemize}
\item $t\bar{t}+3j$;
\item $t\bar{t}+W/Z+2j$;
\item $t\bar{t}t\bar{t}$.
\end{itemize}
with a cut of $H_T>1.5\TeV$ imposed at the generator level to improve reconstruction efficiency.
Even with this selection,
we are computationally limited to processes with at most five final state particles, and restrict ourselves to sufficiently inclusive
quantities in our analysis such that this limitation is unlikely to be important.  
We incorporate the possibility of ``lepton charge flip" manually according to the prescription in Ref~\cite{Aaboud:2017qph}.

\begin{figure}[!hbt]
\centering
\includegraphics[width=0.462\textwidth]{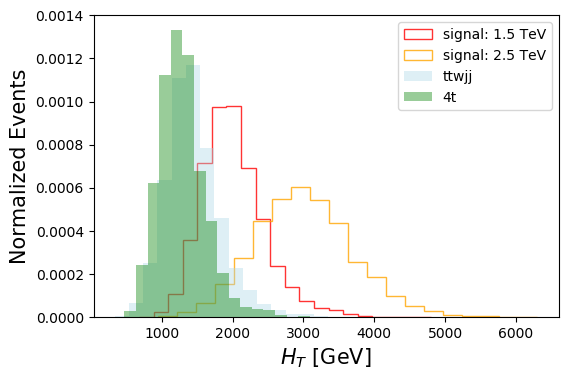}
\includegraphics[width=0.438\textwidth]{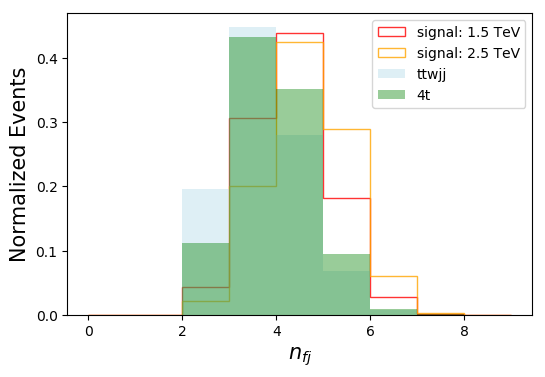}\\
\includegraphics[width=0.462\textwidth]{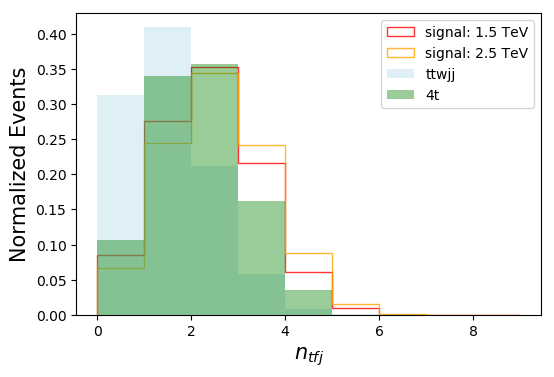}
\includegraphics[width=0.462\textwidth]{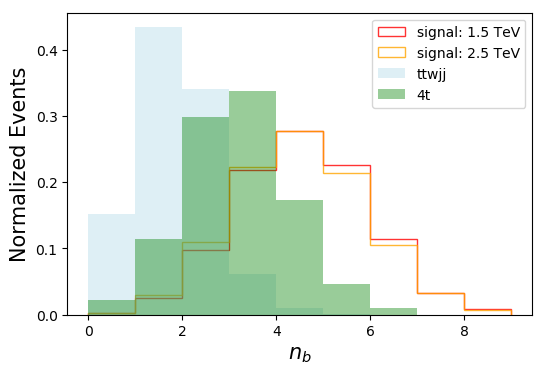}
\caption{The distribution of $H_T$ (top left), $n_{fj}$ (top right), $n_{tfj}$ (bottom left) and $n_b$ (bottom right) of the signal and backgrounds for the 3 mixed sign leptons case. 
Two choices of the top partner mass, $m_{t'}$, are presented: 1.5 TeV (red line) and 2.5 TeV (orange line).
}
\label{fig:AllDistributions}
\end{figure}

After the detector simulation, physics-level objects are reconstructed in both signal and background processes as:
\begin{itemize}
\item Leptons are required to be isolated according to the prescription in Ref.~\cite{Khachatryan:2016yzq}.
\item Jets are reconstructed using the anti-$k_T$ algorithm~\cite{Cacciari:2008gp}
with $r=0.4$ and $p_T > 30$ GeV;
\item Fat jets are reconstructed using anti-$k_T$ with $r=1.0$ and $p_T>200$ GeV;
\item Jets are bottom-tagged according to the DeepFlavor performance shown in Ref.~\cite{CMS-DP-2017-013} using the 70\% tagging efficiency as the work point;
\item Tops are tagged using a convolutional neural network (CNN) described in Appendix.~\ref{app:BJT}
at the 50\% benchmark operating point.
\end{itemize}
These reconstructed objects are fed into the selection described below to assess how well the signal may be
extracted from the background. The distributions of $H_T$, $n_{fj}$ (number of fat jets), $n_{tfj}$ (number of top-tagged jets) and $n_b$ (number of b-tagged jets) from the SM background and the signal
(with two choices of top partner mass, 1.5 TeV (red line) and 2.5 TeV (orange line)) are shown 
in~\autoref{fig:AllDistributions} for the 3 mixed sign leptons case. We can clearly see from this figure that $H_T$ 
of the signal process is usually larger than the background processes and will increase with the mass of the top 
partner, $m_{t'}$. The same behavior also appears in the distributions of $n_{fj}$ and $n_{tfj}$, 
as the more boosted jet is easier to be reconstructed as fat jets and further identified as top jets. The last distribution of $n_b$ is almost 
independent of $m_{t'}$, as it is almost controlled by the true number of the $b$-jets in the 
events, and we model the $b$-tagging efficiency as a constant (70\% as described above) throughout the central region.

\subsection{Event Selection and Sensitivity}

We sort our events into five channels based on the number (and charge) of the leptons they contain as
described above. The event fractions for each channel considering the detector effects are also listed in the third column of Table.~\ref{tab:signal_channel}. Note that we also include 1\% lepton fake rate from jets which results in more leptons than expected just from the branch fraction due to the large multiplicity of jets in the events. For channels with two or more leptons, we eliminate $\mid m_{\ell\ell} - m_Z \mid < 5~$GeV to reduce
background from the $Z$ pole.
At this Pre-Cut selection level, we also require $H_T\geq 2000\GeV$.

After the Pre-Cuts, for $m_{t'}=1.5$~TeV, the signal of 1-$\ell$ and 2-$\ell$ channel is typically 10-100 times 
smaller than the sum of the backgrounds, while other channels have similar with or even larger 
signal than the backgrounds. We further optimize the significance of the top partner signal by 
considering following kinematic variables (Cut I):
\begin{itemize}
\item The number of fat jets $n_{fj} \geq 3$;
\item The number of top tagged fat jets $n_{tfj} \geq 1$;
\item The number of $b$-tagged jets $n_b \geq 5$.
\end{itemize} 
It is likely that the number of untagged jets, $n_j$ is also a useful discriminant.  However, the simulations are limited to five final state particles, $n_j$ may not be modeled well in our simulations, and we do not consider it here. Including this with sophisticated analysis will improve the sensitivity.
For each channel,
the cross section of the signal (for $m_{t'} = 1.5$~TeV)
and corresponding backgrounds after each set of cuts, and the statistical significance of that
channel (assuming 3 ab$^{-1}$ of integrated luminosity)
are summarized in~\tab{cut_flow}.  We find that the single best channel is the one demanding two same sign charged
leptons, which balances rate against standing out from the background.  

\begin{table}[t]
	\begin{center}
		\begin{tabular}{|c|c|c|c|c|c|c|}
		\hline
		Channels & Process & Pre Cut [fb]  & Cut I [fb] &Significance [$3 \text{ab}^{-1}$] \\ 
		\hline
		\multirow{3}{*}{1-$\ell$} & signal & $6.60\times10^{-1}$ & $2.80\times10^{-1}$ & \multirow{3}{*}{20.47 }\\ 
		\cline{2-4}
		&$t_{\ell}t_qjjj$ & $8.28\times10^{1}$ & $4.72\times10^{-1}$ &  \\
		\cline{2-4}
		&$t\bar{t}t\bar{t}$ & $3.97\times10^{-2}$ &$3.27\times10^{-3}$ & \\
		\cline{1-5}
		\multirow{4}{*}{2-os$\ell$} & signal & $1.40\times10^{-1}$ & $5.36\times10^{-2}$ & \multirow{4}{*}{17.99 } \\
		\cline{2-4}
		&$t_{\ell}t_{\ell}jjj$ & $4.71\times10^{0}$ &$1.32\times10^{-2}$& \\
		\cline{2-4}
		&$t_{\ell}t_qW_{\ell}jj$ & $3.01\times10^{-1}$ &$7.70\times10^{-5}$ & \\
		\cline{2-4}
		&$t\bar{t}t\bar{t}$ &$4.14\times10^{-3}$ &$2.54\times10^{-4}$ & \\
		\cline{1-5}
		\multirow{3}{*}{3-ms$\ell$} & signal &$4.30\times10^{-2}$ &$1.45\times10^{-2}$ & \multirow{3}{*}{ 21.41 } \\
		\cline{2-4}
		&$t_{\ell}t_{\ell}W_{\ell}jj$ &$5.30\times10^{-3}$ &$5.89\times10^{-6}$ & \\
		\cline{2-4}
		&$t\bar{t}t\bar{t}$ &$4.36\times10^{-4}$ &$2.25\times10^{-5}$ & \\
		\cline{1-5}
		\multirow{3}{*}{2-ss$\ell$} & signal &$9.57\times10^{-2}$ &$3.67\times10^{-2}$ & \multirow{3}{*}{30.65 } \\
		\cline{2-4}
		&$t_{\ell}t_qW_{\ell}jj$ &$2.91\times10^{-2}$ &$6.48\times10^{-5}$ & \\
		\cline{2-4}
		&$t\bar{t}t\bar{t}$ &$2.11\times10^{-3}$ &$1.31\times10^{-4}$ & \\
		\cline{1-5}
		\multirow{3}{*}{3-ss$\ell$} & signal & $7.16\times10^{-3}$ &/ &\multirow{3}{*}{11.48 } \\
		\cline{2-4}
		&$t_{\ell}t_qW_{\ell}jj$ &$6.45\times10^{-5}$ &/ & \\
		\cline{2-4}
		&$t_{\ell}t_qZ_{\ell\ell}$ & $7.05\times10^{-5}$ &/ & \\
		\hline
		\multicolumn{4}{|r}{Total Significance: } & 47.69 \\
		\hline
\end{tabular}
\caption{Cut flow for $m_{t^{'}}=1.5\TeV$ of all five channels with different number of leptons. The corresponding significance with 3 ab$^{-1}$ luminosity for different channels and the combined significance are also list in the last column. Note that for 3-ss$\ell$ channel, we do not apply Cut I, as the event rate is already extremely low, further selection will decrease the sensitivity. 
}\label{tab:cut_flow}
\end{center}
\end{table}

\begin{figure}
\centering
\includegraphics[width=0.6\textwidth]{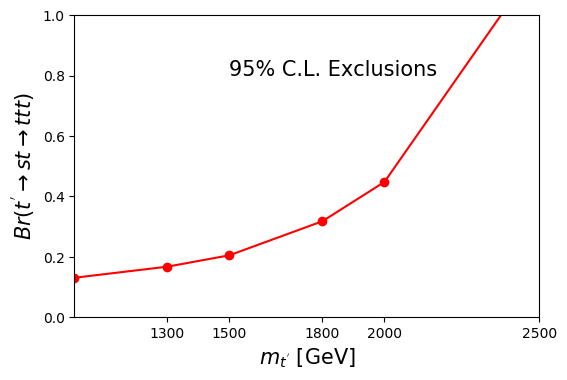}
\caption{The 95\% C.L. upper limit with $L = 3000$ fb$^{-1}$ on the branch fraction of $t'\to st\to t\bar{t}t$ as a function of top-partner mass $m_{t'}$.}
\label{fig:BRlimit}
\end{figure}

For each value of $m_{t'}$, we repeat this procedure for the same set of cuts.  In each case, assuming that the top partners are
pair produced exclusively through the strong force, the sensitivity maps into a bound on the branching ratio for 
$t' \rightarrow t s \rightarrow t \bar{t} t$.
In \autoref{fig:BRlimit}, we show the limit on this branching ratio as a function of $m_{t'}$ from 1000 GeV to 2500 GeV. 
As $m_{t'}$ approaches $2500$ GeV, the upper limit on the branching ratio approaches 1,
implying that higher masses will only be accessible if there is an additional mechanism responsible for producing
$t' \bar{t}'$ beyond the strong interaction.

\subsection{Reconstructing $m_{t'}$}

In the case that an excess is detected, it would be desirable to reconstruct the origin of the signal from top partner
pair production, and determine the $t'$ mass.  Direct reconstruction as an invariant mass is challenging, since
the leptonic top decays produce undetectable neutrino which results in missing momentum, and the decay products of six top quarks result in a large
combinatoric confusion.  

In order to improve the sensitivity to the mass, another CNN is trained to predict the probability 
that a set of events 
originate from a particular value of $m_{t'}$. This CNN has similar structure as the one explained in Appendix.~\ref{app:BJT}. However, instead of the data associated with one particular jet, the whole $p_T$ distribution in the calorimeter for the event after converting into ``tensor image'' is used as the input of the CNN. Using the whole $p_T$ distribution in one event actually captures following two features:
\begin{itemize}
\item The $H_T$ distribution, the sum of the $p_T$ of all visible particles, which increases with $m_{t'}$;
\item The dispersion, which describes the $p_T$ distribution in the whole space, which decreases with $m_{t'}$. 
\end{itemize}

\begin{figure}
\includegraphics[scale=0.33]{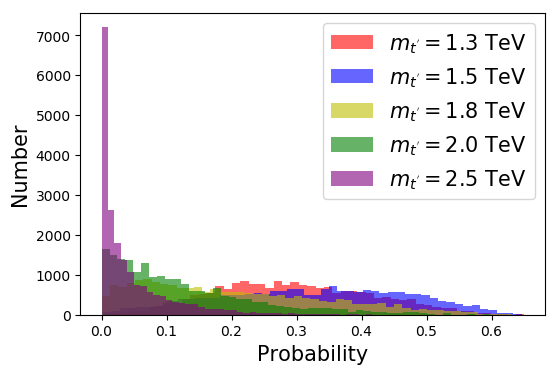}
\includegraphics[scale=0.33]{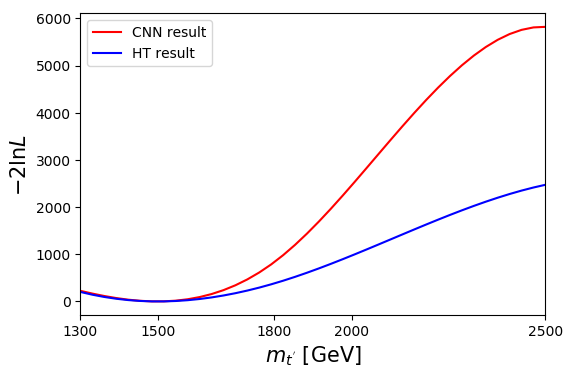}
\includegraphics[scale=0.33]{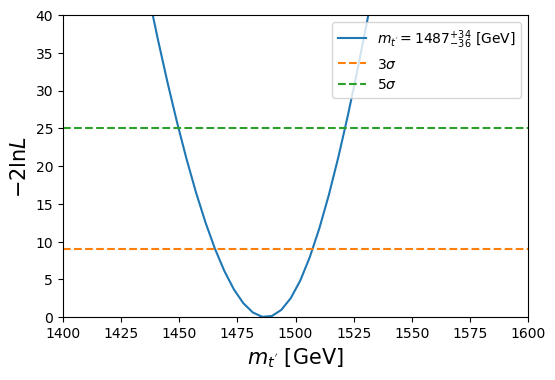}
\caption{Left panel: The CNN output for the 1.5 TeV classifier when fed events corresponding to different values of
$m_{t'}$ as indicated.
Middle  panel: The $-2\ln L$ constructed from the distribution from CNN output (red) and the $H_T$ distribution(blue) aimed for $m_{t'}=1.5$ TeV as a function of the hypothesized mass. Right panel: The $-2\ln L$ analysis around $m_{t^{'}}\sim1.5\TeV$.}
\label{fig:likelihood1}
\end{figure}

We show the output distribution for the 1.5 TeV classifier when
fed simulated events with a variety of values of $m_{t'}$ in the left panel of~\autoref{fig:likelihood1}.
For simplicity, we neglect the background in this assessment; while this is not a good approximation for all of the channels,
it well approximates the channels with the largest sensitivity (such as 2-ss$\ell$).  We leave a more realistic analysis for
future work.

Based on the distributions shown in the left panel of~\autoref{fig:likelihood1}, a binned likelihood is constructed
and its negative log-likelihood is shown in the middle panel of~\autoref{fig:likelihood1}.  Also for comparison, 
the result corresponding to the $H_T$ distribution alone is also presented, illustrating the increase
in sensitivity achieved by the CNN. A more detailed analysis for 1.5 TeV case is shown in the 
right panel of~\autoref{fig:likelihood1}, and an $\mathcal{O}(100)$ GeV determination of the top partner mass can be achieved.

\section{Conclusions}
\label{sec:conclusion}

Events containing six top quarks are within grasp of the LHC Run 3, and provide a fascinating laboratory to search
for physics beyond the Standard Model.  We have explored a simplified model which arises as the low energy limit
of compelling theories of a composite Higgs, and in which top partners decay into three top quarks with a large
branching ratio.  We have constructed inclusive observables which are able to tease the signal out of the otherwise
large Standard Model background, and find that top partner masses up to around 2.5 TeV are accessible with
$\sim 3$~ab$^{-1}$ as can be seen from~\autoref{fig:BRlimit}. 

Further, the distribution of the final state particles also 
provides information about the mass of the top partner. A CNN-based method is used to investigate how well one 
can determine the top partner mass, with the whole $p_T$ distribution over the calorimeter used as the input to the CNN. 
As shown in~\autoref{fig:likelihood1}, around 1.5 TeV, an $\mathcal{O}(100)$ GeV determination of the mass can be achieved.

\begin{acknowledgments}

Y.W. is supported by the Natural Sciences and Engineering Research Council of Canada (NSERC). T.M.P.T. is supported in part by the US National Science Foundation through NSF Grant No.~PHY-1620638. J.S. is supported by the National Natural Science Foundation of China (NSFC) under grant No.11647601, No.11690022, No.11851302, No.11675243 and No.11761141011 and also supported by the Strategic Priority Research Program of the Chinese Academy of Sciences under grant No.XDB21010200 and No.XDB23000000. H.H. is supported by NSFC under grant No.~11847151. T.M. is supported in part by project Y6Y2581B11 supported by 2016 National Postdoctoral Program for Innovative Talents. The simulations for this work were done in part at the HPC Cluster of ITP-CAS.

\end{acknowledgments}

\begin{appendix}
\section{Boosted Jet Tagging}
\label{app:BJT}

Our jet classification is based on a Convolutional Neutral Network (CNN) which combines calorimeter and tracker information for each 
fat jet to assign the probabilities that the jet originates from a top, $W$ boson or light parton. 
For recent work on related strategies, 
see Refs.~\cite{Plehn:2011tg,Kasieczka:2017nvn,Butter:2017cot,Dasgupta:2018emf,Macaluso:2018tck,Csaki:2018hyw}.

The training and testing samples are generated through the same procedure as for the signal and background events, 
simulating the processes $pp \to XX$ with $X=j,\,t$ and $W$.
After reconstructing the fat jets using the anti-$k_t$ algorithm 
with $\Delta R = 1.0$ and $p_T > 200$ GeV, each of them is converted into a ``tensor image''.
A square region in the $(\eta,\phi)$ plane of size $1.0\times1.0$ is constructed
centered at the center of the jet and divided into $50\times50$ equal-sized pixels.
Each pixel records the total incident $p_T$ and the multiplicities of both the track and tower classes (from {\tt Delphes}).
This results in a four channel image with dimensions $50\times50\times4$.

The tensor image serves as the input to the CNN constructed using the {\tt PyTorch} framework. 
The CNN consists of the following elements:
\begin{itemize}
\item Four convolutional layers with a Rectified Linear Unit (ReLU) activation function; 
\item Two \emph{max-pooling} layers;
\item Classification block layers, including two linear layers with a dropout of $50\%$ probability and ReLU activation function;
\item Final linear layer classifying the jet images into different categories.
\end{itemize}

\begin{figure}[!t]
\centering
\includegraphics[width=0.45\textwidth]{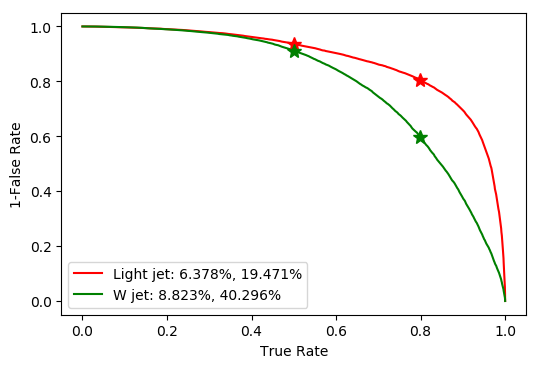}
\includegraphics[width=0.45\textwidth]{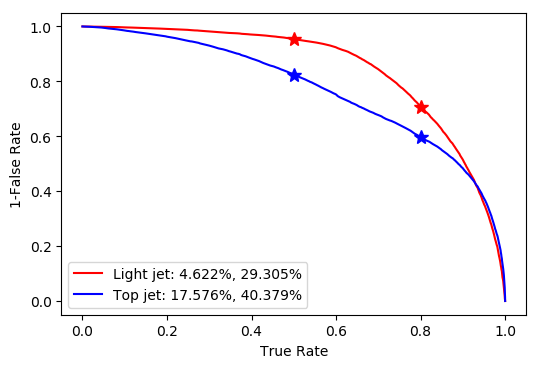}\\
\includegraphics[width=0.45\textwidth]{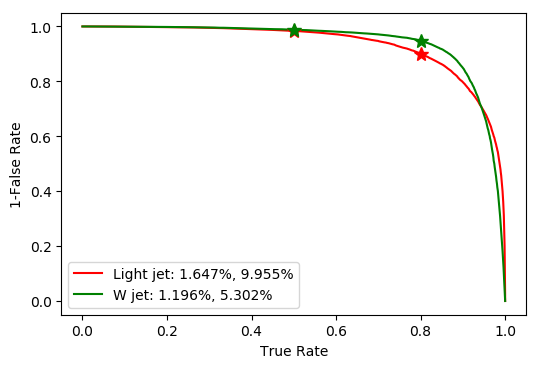}
\includegraphics[width=0.45\textwidth]{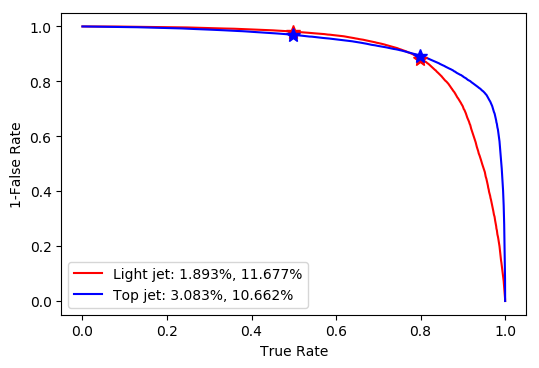}\\
\includegraphics[width=0.45\textwidth]{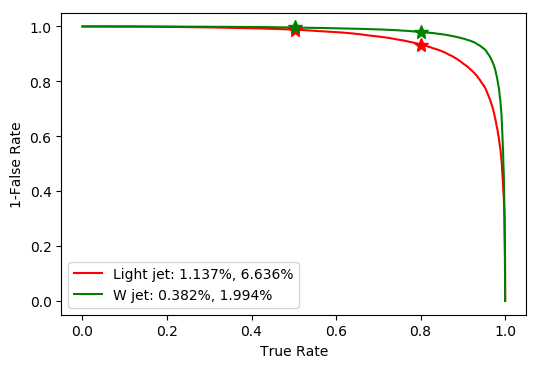}
\includegraphics[width=0.45\textwidth]{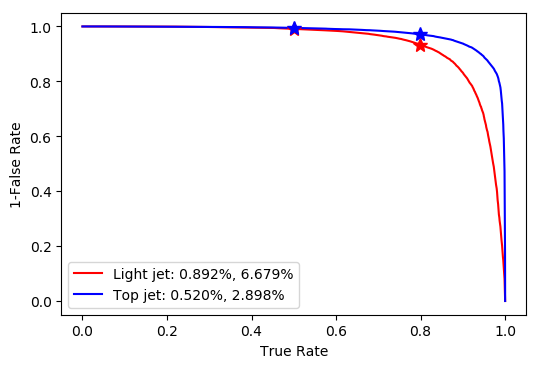}
\caption{The ROC curves for tagging a top (left column) against
a light jet (red) or $W$ (green); or for tagging a $W$ boson (right column) against
a light jet (red) or top (blue), for $p_T$ ranges are
[200, 400] GeV, [400, 800] GeV and $>800$ GeV for the upper, middle and lower rows.
Benchmark points with top tagging rates of 0.5 and 0.8 are indicated on 
each curve, with the corresponding 1-``mistagging rate'' listed in the legend.  }
\label{fig:roc_top_w}
\end{figure}

In each sample, jets are divided into three bins according to their $p_T$: 
200 GeV $< p_T^{\text{jet}}<$ 400 GeV, 
400 GeV $< p_T^{\text{jet}}<$ 800 GeV 
and $p_T^{\text{jet}}>$ 800 GeV, and
the CNN is trained separately for each $p_T$ bin. The tagging performance is characterized by the
\emph{Receiver Operating Characteristic} (ROC) curve. 
For each pair of jet classes $j_1$ and $j_2$ (tagging $j_1$ against $j_2$), 
the ROC curve (see~\autoref{fig:roc_top_w}) shows the ``tagging efficiency'' 
(the probability of correctly tagging the jet of class $j_1$ as $j_1$) on the horizontal axis, 
and 1-``mistagging rate'' (the probability of incorrectly tagging jet of class $j_2$ as $j_1$) on the vertical axis. 

In~\autoref{fig:roc_top_w},
the left panels show the ROC curves for tagging a top quark against a $W$-boson and a light jet, 
while the right panels are the ROC curves for tagging a $W$ boson against a top quark and a light jet.
The top, middle and bottom panels correspond to the $p_T$ bins:  [200,400] GeV, [400,800] GeV 
and [800,$\infty$] GeV, respectively. 
As expected, higher $p_T$ tops and $W$s are identified much more efficiently.
Two benchmark working points corresponding to 50\% and 80\% efficiency for top tagging are marked on each curve in~\autoref{fig:roc_top_w}, 
and the corresponding mistagging rates are listed in the legend of each panel. In practice, the 50\% working point is used to tag the top jets.

\end{appendix}

\clearpage

\bibliographystyle{JHEP}
\bibliography{references}

\providecommand{\href}[2]{#2}\begingroup\raggedright\begin{thebibliography}{10}

\bibitem{Feng:2013pwa}
J.~L. Feng, \emph{{Naturalness and the Status of Supersymmetry}},
  \href{https://doi.org/10.1146/annurev-nucl-102010-130447}{\emph{Ann. Rev.
  Nucl. Part. Sci.} {\bfseries 63} (2013) 351}
  [\href{https://arxiv.org/abs/1302.6587}{{\ttfamily 1302.6587}}].

\bibitem{Giudice:2013nak}
G.~F. Giudice, \emph{{Naturalness after LHC8}}, {\emph{PoS} {\bfseries
  EPS-HEP2013} (2013) 163} [\href{https://arxiv.org/abs/1307.7879}{{\ttfamily
  1307.7879}}].

\bibitem{Altarelli:2013lla}
G.~Altarelli, \emph{{The Higgs: so simple yet so unnatural}},
  \href{https://doi.org/10.1088/0031-8949/2013/T158/014011}{\emph{Phys.
  Scripta} {\bfseries T158} (2013) 014011}
  [\href{https://arxiv.org/abs/1308.0545}{{\ttfamily 1308.0545}}].

\bibitem{Farina:2013mla}
M.~Farina, D.~Pappadopulo and A.~Strumia, \emph{{A modified naturalness
  principle and its experimental tests}},
  \href{https://doi.org/10.1007/JHEP08(2013)022}{\emph{JHEP} {\bfseries 08}
  (2013) 022} [\href{https://arxiv.org/abs/1303.7244}{{\ttfamily 1303.7244}}].

\bibitem{deGouvea:2014xba}
A.~de~Gouvea, D.~Hernandez and T.~M.~P. Tait, \emph{{Criteria for Natural
  Hierarchies}}, \href{https://doi.org/10.1103/PhysRevD.89.115005}{\emph{Phys.
  Rev.} {\bfseries D89} (2014) 115005}
  [\href{https://arxiv.org/abs/1402.2658}{{\ttfamily 1402.2658}}].

\bibitem{Csaki:2018hyw}
C.~Csáki, F.~Ferreira De~Freitas, L.~Huang, T.~Ma, M.~Perelstein and J.~Shu,
  \emph{{Naturalness Sum Rules and Their Collider Tests}},
  \href{https://arxiv.org/abs/1811.01961}{{\ttfamily 1811.01961}}.

\bibitem{Chen:2017dwb}
C.-R. Chen, J.~Hajer, T.~Liu, I.~Low and H.~Zhang, \emph{{Testing naturalness
  at 100 TeV}}, \href{https://doi.org/10.1007/JHEP09(2017)129}{\emph{JHEP}
  {\bfseries 09} (2017) 129}
  [\href{https://arxiv.org/abs/1705.07743}{{\ttfamily 1705.07743}}].

\bibitem{Lillie:2007hd}
B.~Lillie, J.~Shu and T.~M.~P. Tait, \emph{{Top Compositeness at the Tevatron
  and LHC}}, \href{https://doi.org/10.1088/1126-6708/2008/04/087}{\emph{JHEP}
  {\bfseries 04} (2008) 087} [\href{https://arxiv.org/abs/0712.3057}{{\ttfamily
  0712.3057}}].

\bibitem{Pomarol:2008bh}
A.~Pomarol and J.~Serra, \emph{{Top Quark Compositeness: Feasibility and
  Implications}}, \href{https://doi.org/10.1103/PhysRevD.78.074026}{\emph{Phys.
  Rev.} {\bfseries D78} (2008) 074026}
  [\href{https://arxiv.org/abs/0806.3247}{{\ttfamily 0806.3247}}].

\bibitem{Chen:2008hh}
C.-R. Chen, W.~Klemm, V.~Rentala and K.~Wang, \emph{{Color Sextet Scalars at
  the CERN Large Hadron Collider}},
  \href{https://doi.org/10.1103/PhysRevD.79.054002}{\emph{Phys. Rev.}
  {\bfseries D79} (2009) 054002}
  [\href{https://arxiv.org/abs/0811.2105}{{\ttfamily 0811.2105}}].

\bibitem{Kumar:2009vs}
K.~Kumar, T.~M.~P. Tait and R.~Vega-Morales, \emph{{Manifestations of Top
  Compositeness at Colliders}},
  \href{https://doi.org/10.1088/1126-6708/2009/05/022}{\emph{JHEP} {\bfseries
  05} (2009) 022} [\href{https://arxiv.org/abs/0901.3808}{{\ttfamily
  0901.3808}}].

\bibitem{Gregoire:2011ka}
T.~Gregoire, E.~Katz and V.~Sanz, \emph{{Four top quarks in extensions of the
  standard model}},
  \href{https://doi.org/10.1103/PhysRevD.85.055024}{\emph{Phys. Rev.}
  {\bfseries D85} (2012) 055024}
  [\href{https://arxiv.org/abs/1101.1294}{{\ttfamily 1101.1294}}].

\bibitem{Sirunyan:2017roi}
{\scshape CMS} collaboration, A.~M. Sirunyan et~al., \emph{{Search for standard
  model production of four top quarks with same-sign and multilepton final
  states in proton?proton collisions at $\sqrt{s} = 13\,\text {TeV} $}},
  \href{https://doi.org/10.1140/epjc/s10052-018-5607-5}{\emph{Eur. Phys. J.}
  {\bfseries C78} (2018) 140}
  [\href{https://arxiv.org/abs/1710.10614}{{\ttfamily 1710.10614}}].

\bibitem{Aaboud:2018jsj}
{\scshape ATLAS} collaboration, M.~Aaboud et~al., \emph{{Search for
  four-top-quark production in the single-lepton and opposite-sign dilepton
  final states in pp collisions at $\sqrt{s}$ = 13 TeV with the ATLAS
  detector}}, {\emph{Submitted to: Phys. Rev.} (2018) }
  [\href{https://arxiv.org/abs/1811.02305}{{\ttfamily 1811.02305}}].

\bibitem{Csaki:2017jby}
C.~Csáki, T.~Ma and J.~Shu, \emph{{Trigonometric Parity for the Composite
  Higgs}}, \href{https://doi.org/10.1103/PhysRevLett.121.231801}{\emph{Phys.
  Rev. Lett.} {\bfseries 121} (2018) 231801}
  [\href{https://arxiv.org/abs/1709.08636}{{\ttfamily 1709.08636}}].

\bibitem{Ryttov:2008xe}
T.~A. Ryttov and F.~Sannino, \emph{{Ultra Minimal Technicolor and its Dark
  Matter TIMP}}, \href{https://doi.org/10.1103/PhysRevD.78.115010}{\emph{Phys.
  Rev.} {\bfseries D78} (2008) 115010}
  [\href{https://arxiv.org/abs/0809.0713}{{\ttfamily 0809.0713}}].

\bibitem{Galloway:2010bp}
J.~Galloway, J.~A. Evans, M.~A. Luty and R.~A. Tacchi, \emph{{Minimal Conformal
  Technicolor and Precision Electroweak Tests}},
  \href{https://doi.org/10.1007/JHEP10(2010)086}{\emph{JHEP} {\bfseries 10}
  (2010) 086} [\href{https://arxiv.org/abs/1001.1361}{{\ttfamily 1001.1361}}].

\bibitem{Gripaios:2009pe}
B.~Gripaios, A.~Pomarol, F.~Riva and J.~Serra, \emph{{Beyond the Minimal
  Composite Higgs Model}},
  \href{https://doi.org/10.1088/1126-6708/2009/04/070}{\emph{JHEP} {\bfseries
  04} (2009) 070} [\href{https://arxiv.org/abs/0902.1483}{{\ttfamily
  0902.1483}}].

\bibitem{Frigerio:2012uc}
M.~Frigerio, A.~Pomarol, F.~Riva and A.~Urbano, \emph{{Composite Scalar Dark
  Matter}}, \href{https://doi.org/10.1007/JHEP07(2012)015}{\emph{JHEP}
  {\bfseries 07} (2012) 015} [\href{https://arxiv.org/abs/1204.2808}{{\ttfamily
  1204.2808}}].

\bibitem{Serra:2017poj}
J.~Serra and R.~Torre, \emph{{Neutral naturalness from the brother-Higgs
  model}}, \href{https://doi.org/10.1103/PhysRevD.97.035017}{\emph{Phys. Rev.}
  {\bfseries D97} (2018) 035017}
  [\href{https://arxiv.org/abs/1709.05399}{{\ttfamily 1709.05399}}].

\bibitem{Ma:2015gra}
T.~Ma and G.~Cacciapaglia, \emph{{Fundamental Composite 2HDM: SU(N) with 4
  flavours}}, \href{https://doi.org/10.1007/JHEP03(2016)211}{\emph{JHEP}
  {\bfseries 03} (2016) 211}
  [\href{https://arxiv.org/abs/1508.07014}{{\ttfamily 1508.07014}}].

\bibitem{Ma:2017vzm}
Y.~Wu, T.~Ma, B.~Zhang and G.~Cacciapaglia, \emph{{Composite Dark Matter and
  Higgs}}, \href{https://doi.org/10.1007/JHEP11(2017)058}{\emph{JHEP}
  {\bfseries 11} (2017) 058}
  [\href{https://arxiv.org/abs/1703.06903}{{\ttfamily 1703.06903}}].

\bibitem{Cacciapaglia:2018avr}
G.~Cacciapaglia, S.~Vatani, T.~Ma and Y.~Wu, \emph{{Towards a fundamental safe
  theory of composite Higgs and Dark Matter}},
  \href{https://arxiv.org/abs/1812.04005}{{\ttfamily 1812.04005}}.

\bibitem{Aad:2016tuk}
{\scshape ATLAS} collaboration, G.~Aad et~al., \emph{{Search for supersymmetry
  at $\sqrt{s}=13$ TeV in final states with jets and two same-sign leptons or
  three leptons with the ATLAS detector}},
  \href{https://doi.org/10.1140/epjc/s10052-016-4095-8}{\emph{Eur. Phys. J.}
  {\bfseries C76} (2016) 259}
  [\href{https://arxiv.org/abs/1602.09058}{{\ttfamily 1602.09058}}].

\bibitem{ATLAS-CONF-2016-013}
\emph{{Search for production of vector-like top quark pairs and of four top
  quarks in the lepton-plus-jets final state in $pp$ collisions at
  $\sqrt{s}=13$ TeV with the ATLAS detector}},  Tech. Rep. ATLAS-CONF-2016-013,
  CERN, Geneva, Mar, 2016.

\bibitem{Aaboud:2017dmy}
{\scshape ATLAS} collaboration, M.~Aaboud et~al., \emph{{Search for
  supersymmetry in final states with two same-sign or three leptons and jets
  using 36 fb$^{-1}$ of $\sqrt{s}=13$ TeV $pp$ collision data with the ATLAS
  detector}}, \href{https://doi.org/10.1007/JHEP09(2017)084}{\emph{JHEP}
  {\bfseries 09} (2017) 084}
  [\href{https://arxiv.org/abs/1706.03731}{{\ttfamily 1706.03731}}].

\bibitem{Aaboud:2018zeb}
{\scshape ATLAS} collaboration, M.~Aaboud et~al., \emph{{Search for
  supersymmetry in events with four or more leptons in $\sqrt{s}=13$ TeV $pp$
  collisions with ATLAS}},
  \href{https://doi.org/10.1103/PhysRevD.98.032009}{\emph{Phys. Rev.}
  {\bfseries D98} (2018) 032009}
  [\href{https://arxiv.org/abs/1804.03602}{{\ttfamily 1804.03602}}].

\bibitem{Sirunyan:2017lae}
{\scshape CMS} collaboration, A.~M. Sirunyan et~al., \emph{{Search for
  electroweak production of charginos and neutralinos in multilepton final
  states in proton-proton collisions at $\sqrt{s}=$ 13 TeV}},
  \href{https://doi.org/10.1007/JHEP03(2018)166}{\emph{JHEP} {\bfseries 03}
  (2018) 166} [\href{https://arxiv.org/abs/1709.05406}{{\ttfamily
  1709.05406}}].

\bibitem{Khachatryan:2017qgo}
{\scshape CMS} collaboration, V.~Khachatryan et~al., \emph{{Search for new
  phenomena with multiple charged leptons in proton–proton collisions at
  $\sqrt{s}= 13$ $\,\text {TeV}$}},
  \href{https://doi.org/10.1140/epjc/s10052-017-5182-1}{\emph{Eur. Phys. J.}
  {\bfseries C77} (2017) 635}
  [\href{https://arxiv.org/abs/1701.06940}{{\ttfamily 1701.06940}}].

\bibitem{Aad:2015fna}
{\scshape ATLAS} collaboration, G.~Aad et~al., \emph{{A search for $
  t\overline{t} $ resonances using lepton-plus-jets events in proton-proton
  collisions at $ \sqrt{s}=8 $ TeV with the ATLAS detector}},
  \href{https://doi.org/10.1007/JHEP08(2015)148}{\emph{JHEP} {\bfseries 08}
  (2015) 148} [\href{https://arxiv.org/abs/1505.07018}{{\ttfamily
  1505.07018}}].

\bibitem{Aaboud:2017nfd}
{\scshape ATLAS} collaboration, M.~Aaboud et~al., \emph{{Search for direct top
  squark pair production in final states with two leptons in $\sqrt{s} = 13$
  TeV $pp$ collisions with the ATLAS detector}},
  \href{https://doi.org/10.1140/epjc/s10052-017-5445-x}{\emph{Eur. Phys. J.}
  {\bfseries C77} (2017) 898}
  [\href{https://arxiv.org/abs/1708.03247}{{\ttfamily 1708.03247}}].

\bibitem{Aaboud:2017ayj}
{\scshape ATLAS} collaboration, M.~Aaboud et~al., \emph{{Search for a scalar
  partner of the top quark in the jets plus missing transverse momentum final
  state at $\sqrt{s}$=13 TeV with the ATLAS detector}},
  \href{https://doi.org/10.1007/JHEP12(2017)085}{\emph{JHEP} {\bfseries 12}
  (2017) 085} [\href{https://arxiv.org/abs/1709.04183}{{\ttfamily
  1709.04183}}].

\bibitem{Aaboud:2017aeu}
{\scshape ATLAS} collaboration, M.~Aaboud et~al., \emph{{Search for top-squark
  pair production in final states with one lepton, jets, and missing transverse
  momentum using 36 fb$^{?1}$ of $ \sqrt{s}=13 $ TeV pp collision data with the
  ATLAS detector}}, \href{https://doi.org/10.1007/JHEP06(2018)108}{\emph{JHEP}
  {\bfseries 06} (2018) 108}
  [\href{https://arxiv.org/abs/1711.11520}{{\ttfamily 1711.11520}}].

\bibitem{Drees:2013wra}
M.~Drees, H.~Dreiner, D.~Schmeier, J.~Tattersall and J.~S. Kim,
  \emph{{CheckMATE: Confronting your Favourite New Physics Model with LHC
  Data}}, \href{https://doi.org/10.1016/j.cpc.2014.10.018}{\emph{Comput. Phys.
  Commun.} {\bfseries 187} (2015) 227}
  [\href{https://arxiv.org/abs/1312.2591}{{\ttfamily 1312.2591}}].

\bibitem{Aaboud:2018mjh}
{\scshape ATLAS} collaboration, M.~Aaboud et~al., \emph{{Search for heavy
  particles decaying into top-quark pairs using lepton-plus-jets events in
  proton–proton collisions at $\sqrt{s} = 13$ $\text {TeV}$ with the ATLAS
  detector}}, \href{https://doi.org/10.1140/epjc/s10052-018-5995-6}{\emph{Eur.
  Phys. J.} {\bfseries C78} (2018) 565}
  [\href{https://arxiv.org/abs/1804.10823}{{\ttfamily 1804.10823}}].

\bibitem{Alwall:2014hca}
J.~Alwall, R.~Frederix, S.~Frixione, V.~Hirschi, F.~Maltoni, O.~Mattelaer
  et~al., \emph{{The automated computation of tree-level and next-to-leading
  order differential cross sections, and their matching to parton shower
  simulations}}, \href{https://doi.org/10.1007/JHEP07(2014)079}{\emph{JHEP}
  {\bfseries 07} (2014) 079} [\href{https://arxiv.org/abs/1405.0301}{{\ttfamily
  1405.0301}}].

\bibitem{Nadolsky:2008zw}
P.~M. Nadolsky, H.-L. Lai, Q.-H. Cao, J.~Huston, J.~Pumplin, D.~Stump et~al.,
  \emph{{Implications of CTEQ global analysis for collider observables}},
  \href{https://doi.org/10.1103/PhysRevD.78.013004}{\emph{Phys. Rev.}
  {\bfseries D78} (2008) 013004}
  [\href{https://arxiv.org/abs/0802.0007}{{\ttfamily 0802.0007}}].

\bibitem{Artoisenet:2012st}
P.~Artoisenet, R.~Frederix, O.~Mattelaer and R.~Rietkerk, \emph{{Automatic
  spin-entangled decays of heavy resonances in Monte Carlo simulations}},
  \href{https://doi.org/10.1007/JHEP03(2013)015}{\emph{JHEP} {\bfseries 03}
  (2013) 015} [\href{https://arxiv.org/abs/1212.3460}{{\ttfamily 1212.3460}}].

\bibitem{Sjostrand:2014zea}
T.~Sjöstrand, S.~Ask, J.~R. Christiansen, R.~Corke, N.~Desai, P.~Ilten et~al.,
  \emph{{An Introduction to PYTHIA 8.2}},
  \href{https://doi.org/10.1016/j.cpc.2015.01.024}{\emph{Comput. Phys. Commun.}
  {\bfseries 191} (2015) 159}
  [\href{https://arxiv.org/abs/1410.3012}{{\ttfamily 1410.3012}}].

\bibitem{deFavereau:2013fsa}
{\scshape DELPHES 3} collaboration, J.~de~Favereau, C.~Delaere, P.~Demin,
  A.~Giammanco, V.~Lemaître, A.~Mertens et~al., \emph{{DELPHES 3, A modular
  framework for fast simulation of a generic collider experiment}},
  \href{https://doi.org/10.1007/JHEP02(2014)057}{\emph{JHEP} {\bfseries 02}
  (2014) 057} [\href{https://arxiv.org/abs/1307.6346}{{\ttfamily 1307.6346}}].

\bibitem{Cacciari:2011ma}
M.~Cacciari, G.~P. Salam and G.~Soyez, \emph{{FastJet User Manual}},
  \href{https://doi.org/10.1140/epjc/s10052-012-1896-2}{\emph{Eur. Phys. J.}
  {\bfseries C72} (2012) 1896}
  [\href{https://arxiv.org/abs/1111.6097}{{\ttfamily 1111.6097}}].

\bibitem{Aaboud:2017qph}
{\scshape ATLAS} collaboration, M.~Aaboud et~al., \emph{{Search for doubly
  charged Higgs boson production in multi-lepton final states with the ATLAS
  detector using proton–proton collisions at $\sqrt{s}=13\,\text {TeV}$}},
  \href{https://doi.org/10.1140/EPJC/S10052-018-5661-Z,
  10.1140/epjc/s10052-018-5661-z}{\emph{Eur. Phys. J.} {\bfseries C78} (2018)
  199} [\href{https://arxiv.org/abs/1710.09748}{{\ttfamily 1710.09748}}].

\bibitem{Khachatryan:2016yzq}
{\scshape CMS} collaboration, V.~Khachatryan et~al., \emph{{Measurements of the
  $\mathrm{t}\overline{\mathrm{t}}$ production cross section in lepton+jets
  final states in pp collisions at 8 $\,\text {TeV}$ and ratio of 8 to 7
  $\,\text {TeV}$ cross sections}},
  \href{https://doi.org/10.1140/epjc/s10052-016-4504-z}{\emph{Eur. Phys. J.}
  {\bfseries C77} (2017) 15}
  [\href{https://arxiv.org/abs/1602.09024}{{\ttfamily 1602.09024}}].

\bibitem{Cacciari:2008gp}
M.~Cacciari, G.~P. Salam and G.~Soyez, \emph{{The Anti-k(t) jet clustering
  algorithm}}, \href{https://doi.org/10.1088/1126-6708/2008/04/063}{\emph{JHEP}
  {\bfseries 04} (2008) 063} [\href{https://arxiv.org/abs/0802.1189}{{\ttfamily
  0802.1189}}].

\bibitem{CMS-DP-2017-013}
{\scshape CMS Collaboration} collaboration, \emph{{CMS Phase 1 heavy flavour
  identification performance and developments}},  Tech. Rep. CMS-DP-2017-013,
  May, 2017.

\bibitem{Plehn:2011tg}
T.~Plehn and M.~Spannowsky, \emph{{Top Tagging}},
  \href{https://doi.org/10.1088/0954-3899/39/8/083001}{\emph{J. Phys.}
  {\bfseries G39} (2012) 083001}
  [\href{https://arxiv.org/abs/1112.4441}{{\ttfamily 1112.4441}}].

\bibitem{Kasieczka:2017nvn}
G.~Kasieczka, T.~Plehn, M.~Russell and T.~Schell, \emph{{Deep-learning Top
  Taggers or The End of QCD?}},
  \href{https://doi.org/10.1007/JHEP05(2017)006}{\emph{JHEP} {\bfseries 05}
  (2017) 006} [\href{https://arxiv.org/abs/1701.08784}{{\ttfamily
  1701.08784}}].

\bibitem{Butter:2017cot}
A.~Butter, G.~Kasieczka, T.~Plehn and M.~Russell, \emph{{Deep-learned Top
  Tagging with a Lorentz Layer}},
  \href{https://doi.org/10.21468/SciPostPhys.5.3.028}{\emph{SciPost Phys.}
  {\bfseries 5} (2018) 028} [\href{https://arxiv.org/abs/1707.08966}{{\ttfamily
  1707.08966}}].

\bibitem{Dasgupta:2018emf}
M.~Dasgupta, M.~Guzzi, J.~Rawling and G.~Soyez, \emph{{Top tagging : an
  analytical perspective}},
  \href{https://doi.org/10.1007/JHEP09(2018)170}{\emph{JHEP} {\bfseries 09}
  (2018) 170} [\href{https://arxiv.org/abs/1807.04767}{{\ttfamily
  1807.04767}}].

\bibitem{Macaluso:2018tck}
S.~Macaluso and D.~Shih, \emph{{Pulling Out All the Tops with Computer Vision
  and Deep Learning}},
  \href{https://doi.org/10.1007/JHEP10(2018)121}{\emph{JHEP} {\bfseries 10}
  (2018) 121} [\href{https://arxiv.org/abs/1803.00107}{{\ttfamily
  1803.00107}}].

\end{thebibliography}\endgroup

\end{document}